# The Future of Work: Agile in a Hybrid World

Dennis Mancl and Steven D. Fraser

**Abstract.** An agile organization adapts what they are building to match their customer's evolving needs. Agile teams also adapt to changes in their organization's work environment. The latest change is the evolving environment of "hybrid" work – a mix of in-person and virtual staff. Team members might sometimes work together in the office, work from home, or work in other locations, and they may struggle to sustain a high level of collaboration and innovation. It isn't just pandemic social distancing – many of us want to work from home to eliminate our commute and spend more time with family. Are there learnings and best practices that organizations can use to become and stay effective in a hybrid world? An XP 2022 panel organized by Steven Fraser (Innoxec) discussed these questions in June 2022. The panel was facilitated by Hendrik Esser (Ericsson) and featured Alistair Cockburn (Heart of Agile), Sandy Mamoli (Nomad8), Nils Brede Moe (SINTEF), Jaana Nyfjord (Spotify), and Darja Smite (Blekinge Institute of Technology).

## 1. Introduction: Panel discussion of agility today

This paper reports on a panel session organized by Steven Fraser (Innoxec) and facilitated by Hendrik Esser (Ericsson) to discuss XP 2022's theme "The Future of Work: Agile in a Hybrid World." The panel was also inspired by a recent community survey by Fraser and Mancl on "The Future of Conferences" [1] in an increasingly virtual world. In preparation for the panel, panelists were asked to consider topics including:

- As pandemic restrictions are loosened, how will work environments and Agile practices evolve, given that some workers will return to a physical office while others may desire to continue their work from home?
- What issues are related to surveillance, trust, collaboration, and networking tools?
- What can we leverage from the "Peopleware" inspired writings of Fred Brooks [2], Melvin Conway [3], and Tom DeMarco and Tim Lister [4][5]?
- What are the key learnings and emergent best practices/hybrid for hybrid work environments for Agile organizations?

In response to the pre-panel questions, Alistair Cockburn (Heart of Agile, USA) raised several concerns:

- When people do "hybrid-agile," have they created the worst of both worlds?
- How has something fundamentally about human interaction become a technology challenge, where tools matter, and tool-usage matters more!
- What are best-practice facilitation skills for effective virtual/hybrid collaboration?
- What is the future of work? How will current company best practices evolve?

At XP 2022, in Copenhagen, the panelists shared their experiences with virtual and hybrid working environments during the pandemic, data from software developer surveys, and key ideas from previous generations of workplace evolution. Agile values and practices were important, however the panel also considered broader issues:

- Do individuals have freedom of choice regarding their own work environment?
- What are the best ways to organize work teams so they can collaborate effectively?
- How can they best approach creative problem-solving as a team?

"Adaptation" is an important part of agility. Agile teams need to adapt to changes in their organization's work environment. The latest change is the evolving environment of "hybrid" work, a mix of in-person and virtual staff. Team members might work in the office, from home, or in other locations, but there will always be a struggle to achieve good collaboration and innovation. The motivation for virtual work isn't just pandemic social distancing: many of us want to work virtually from home to eliminate our commute and spend more time with family.

The panel facilitator, Hendrik Esser (Ericsson, Germany), kicked off the discussion of hybrid working with an interesting anecdote, and his story is partial proof that employees are not necessarily consistent in their beliefs. In Hendrik's office, one person who had resisted a return to the office was mandated to return two days per week. Within one week, his reaction was "I had forgotten how great it is to meet people by the coffee machine, so I think I will come more often." Our personal beliefs about our personal "best" work environment will likely continue to evolve.

Hendrik also noted that the panel would focus more on how to work effectively in a hybrid world – rather than debating the merits of hybrid versus in-person work.

## 2. What is hybrid? Individual choice or team choice?

The panel abstract described "hybrid" work as an evolving environment with a mix of in-person (traditional office) and virtual (remote) work environments.

However, Darja Smite (Blekinge Institute of Technology, Sweden) saw hybrid as two sides of a coin: individual liberty versus potential team conflicts. On one side, hybrid is about flexibility, the freedom to decide where and when to work. She believed that freedom is important. However, she also acknowledged that hybrid can be problematic. Everyone desires freedom of choice regarding workplace, but Darja cautioned, "I see teams with internal conflicts, with members who have different opinions about where they want to work and how they want to collaborate."

Sandy Mamoli (Nomad8, New Zealand) suggested the ambiguity of going hybrid. "We think we can assume that hybrid is here to stay, but we don't know exactly what it is going to look like." She explained that we will need to explore new ways of working, but we must also balance our own preferences against those of the team.

Jaana Nyfjord (Spotify, Sweden) introduced Spotify's perspective on virtual work: "Distributed first" and "work from anywhere." Spotify encourages a non-traditional work structure. The company desires that their staff be as creative and productive as possible and enables staff to choose their workplace structure.

Darja discussed how teams might address the conflicts of work styles. One possible solution is to reorganize teams. "We ask them to imagine reshuffling the people, letting them decide how they want to work. Some teams will be fully off-site teams, some fully on-site teams, and some hybrid teams." It's important "to align team membership with their preferences."

Nils Brede Moe (SINTEF, Norway) agreed with Darja, "There needs to be some compromise." Teams often struggle when people within a single team want so many different things. Nils indicated that experienced team members can leverage an existing network of contacts for assistance, whether they are co-located or virtual. However, new team members may work more effectively face-to-face.

Sandy discussed the consequences of compromise. "I have an unpopular opinion: I don't think it should be 100% individual choice." Sandy explained that she has colleagues who say, "I want to live at the beach, I want to go to the office every Wednesday, and I only think about what's fine for me." They may not think about the bigger picture.

But Sandy believed that a top-down mandate with a single universal rule about virtual working would be wrong. "I don't think it should be dictated by managers. I think there needs to be a conversation, with your peers at least, about what works for us... I want to make sure the whole plan works, and that factors into my decision. It's not just 'me, me, me.'"

## 3. Establishing trust

Alistair Cockburn (Heart of Agile, USA) was skeptical about virtual work. Teams that do not share a physical workspace may find it difficult to establish trust. Alistair suggested that organizations might address this with a "budget line-item for building trust."

This "budget" is about improving the way teams work together. Co-located teams have group activities to support communication and trust within the team: "If you are co-located in the same city, trust is [mostly] for free. [You build trust in the team] when you have a movie night, or maybe a birthday cake. You don't get that when you're distributed." If distributed or hybrid teams are going to build trust, it will cost more: company management may need to budget for travel and lodging expenses and/or multiple team-building events. When a business unit has a budget line-item for trust, the cost of a distributed team becomes more visible to management.

Hendrik agreed about the value of trust: "I think the topic of trust and psychological safety is super-important." The question is "how" to do it in a distributed environment.

## 4. Teamwork, creativity, and execution

Agile practices are designed to reinforce the importance of good communication and collaboration. Some of the practices may even improve morale in distributed teams. Nils pointed this out early in the panel discussion. "Pair programming is a very important practice. Pair programming is what has kept people going during the pandemic, because they have pair programming sessions." Even though everyone worked virtually, pair programming provided a sense of normality and technical interaction with others. "We found that pair programming is even better for some doing it virtually, because you aren't disturbing others."

Darja added more information about this from a recent study [6] interview: "One person explained that his at-home experience depended on the week. He had one week alone at home [not doing any pair programming]. He wasn't taking showers; he was sitting alone on the sofa all day. He was having a miserable work experience, he was not disciplined, and he was very unfocused. Every other week, he was paired with someone – working together for most of the day – so he would dress, he would sit by the desk, he would be motivated to be focusing on work. It was a completely different experience for the same person."

Most of the panelists agreed that face-to-face interaction may be necessary for design discussions and creative problem-solving sessions. Some group activities are much harder to do over a computer connection.

Alistair desired that organizations considering hybrid work, consider how to keep their people happy and productive. He gave some examples of creative face-to-face problem-solving sessions – co-design with two people working together. Most people find it valuable to be involved in some group design work, and in agile development it helps to be able to sit and discuss designs with the customer. However, there needs to be a balance. When people return to the office, "I hope we start to regulate the balance between 'I'm OK working by myself' or 'This is a problem-solving session.'"

Alistair also wondered if quality issues increased during the pandemic. Alistair thought that there has been a general decline in creativity and a related decline in software quality – affecting the usability of web-based user interfaces. As a frequent flyer, Alistair lamented the poor usability of airline websites. "I'm wondering whether in the pandemic, the people designing the UIs are all doing it distributed and they never get their brains together to solve the problems."

Jaana indicated that her company (Spotify) was already proficient with virtual design work prior to the pandemic. When the pandemic hit, Spotify had existing virtual collaboration tooling and processes in place. Teams didn't change much the way they worked.

Sandy responded that the research literature [7] indicates that "we *are* more creative if people are together." We spark off more ideas when we are together because the framing of a face-to-face meeting is more inclusive. We have a richer interaction with others in the room. Nils noted the complex dynamics of a creative meetings: "I think that misunderstandings is the key, that you need to have disagreements to drive the innovation." But teams who have worked together in creative brainstorming sessions *might* be able to be reasonably productive in virtual meetings: "We see that those who made that work before seem to make it work while they are distributed."

But Sandy also defended virtual meetings. It isn't just brainstorming and ideation that steer the creative process, Sandy explained. The same research studies have found that decision making can be better in a virtual meeting. "I found really surprising that decision making was better when people were remote, because there were no distractions – and when they had to do 'dot voting' where people would vote on 'which of those creative ideas to go for,' the results were better when they were distributed."

Alistair suggested that in brainstorming and dot voting, people tend to follow their authority figure. With virtual sessions, voting can be anonymous with "better" results since people are less influenced by hierarchy.

### 5. Why come to the office?

The panelists turned to the audience for their opinions about effective work environments. Comments included:

- Mental health. So I don't go cubby-crazy at home.
- Collaboration. Getting together with the other departments.

- Work style. I work in two completely different ways at home and in the office. At home, I am very focused on tasks and productive, whereas I go to the office because I need frequent coordination with my co-workers.
- I like to work remote since I care for my son and take him to and from school. However, I still go to the office for three or four meetings a month.
- Isolation. There were times when I went to the office and almost all my team was remote, or vice versa. It was terrible and it made me feel isolated.
- Business crisis. Hybrid seems appropriate when everything works, however if your organization struggles, hybrid is an accelerator of disfunction.

Nils also reported on data from his study [8] on "return to the office." "What people want to do has been pretty stable for the last year. Everyone wants flexibility, but everyone wants an office. Younger people want to meet other young people, and they want to do that at the office, at least in our survey."

## 6. Why should I turn on my camera?

An interesting side topic of the panel was webcam etiquette. Alistair started the discussion. He explained that he was recently invited to give a guest lecture at a local university (University of Utah), and that there were in-person and virtual attendees for his talk. He requested that any online attendee with a question should turn on their camera: "Please do me the honor of showing your face."

One person in the audience was Alistair's son, a University of Utah student, who complained, "I don't think you realize how big of an 'ask' that was. I haven't turned on my camera in two years." Alistair was amazed to hear how students were so disconnected from human contact during the pandemic.

Darja went further. She explained that she has seen the same behavior, and not just with young students. "I have spoken to people in companies who are not that young, and they have not turned on their cameras for a number of months and haven't seen their colleagues. I'm not surprised about the younger generation. But I was surprised to see that corporate people with seniority would not turn on their cameras. Not just for someone from a university or a guest lecturer, but for their teammates."

One explanation, from both Darja and Jaana, is that there are companies who have had to work around network bandwidth problems for their online meetings. This was an issue for many companies early in the pandemic. Staff were requested to turn off cameras due to limited bandwidth. Later in the pandemic, after companies had upgraded their networks, everyone still followed old habits.

The panelists agreed that issues of "psychological safety" may motivate people to keep their cameras off. Most do not wish to be "surveilled," eight hours a day. If a meeting is not interesting, it's easier to focus on other tasks when no one can see you. Also – working from home with background commotion, most people prefer to have their cameras off and their microphones muted.

## 7. How to work effectively in the future

Hendrik concluded the panel with one final question. "What would be your advice to companies and managers?"

Jaana suggested that we have an uncertain future, so we should focus on making micro-improvements.

Sandy believed that we should give teams more freedom and flexibility to decide how and where to work, but not necessarily individuals.

Nils advocated for more teamwork. We need to solve tasks together and work together more. It's easier when you are in the office, but it can be made to work in a hybrid environment.

Darja admitted to being excited to see the transformations of the "Workplace of the Future" and how offices will be reinvented.

Alistair gave some simple advice: Start with two days a week in the office and use that time to maximize productivity and trust-building.

## 8. Summary

While the panel offered some practical advice, it reached no firm conclusions. However, it is clear that the pandemic has changed how we approach Agile today. Practices that once required in-person co-location have evolved and met the challenge of virtual networking.

Virtual and hybrid work have also created new opportunities. Technology will play a role in promoting a diverse and inclusive workforce. An increase in workplace access for women, minorities, and the disabled has been described in academic studies [9] and in popular media [10][11]. In the US, federal regulations require employers to make "reasonable workplace accommodations" for disabled workers. Because virtual work has become more established across industry during the pandemic, legal journals have suggested how workers may qualify for mandated virtual or hybrid work arrangements [12]. The case law is still in flux, and different rules may apply in other countries.

A conflict between workers and managers in the post-pandemic drive to "return everyone to the office" has been reported in the popular media. It may be best to follow an Agile approach: be flexible and work together as a team whether in-person, hybrid, or virtual. Agile workgroups see workplace issues that call for team-based solutions, including team members who are reluctant to return to the office [13], uncertainty about the effectiveness of hybrid work [14], and even concerns about new kinds of workplace surveillance [15]. Over time, we expect that organizations will adapt to meet the demands of a hybrid world.

References
1. Fraser, S., Mancl, D.: The Future of Conferences Research Survey, https://manclswx.com/survey2022.html. Accessed 16 Jul. 2022 (2022)
2. Brooks, F.P.: The Mythical Man-Month. Addison-Wesley, Boston, MA (1995)
3. Conway, M.E.: How Do Committees Invent? Datamation (1968)
4. DeMarco, T., Lister, T.: Peopleware, Addison-Wesley, Boston, MA (2013)
5. Fraser, S. et al.: Retrospectives on Peopleware. In: ICSE Companion. pp. 21–24 IEEE Computer Society (2007). https://doi.org/10.1109/ICSECOMPANION.2007.61
6. Smite, D., Moe, N.B., Klotins, E., Gonzalez-Huerta, J.: From Forced Working-From-Home to Working-From-Anywhere: Two Revolutions in Telework. https://arxiv.org/abs/2101.08315. Accessed 22 Jul. 2022 (2021)
7. Brucks, M.S., Levav, J.: Virtual communication curbs creative idea generation. Nature 605, 108–112 (2022). https://doi.org/10.1038/s41586-022-04643-y
8. Smite, D., Moe, N.B., Hildrum, J., Gonzalez Huerta, J., Mendez, D.: Work-From-Home is Here to Stay: Call for Flexibility in Post-Pandemic Work Policies. https://arxiv.org/abs/2203.11136. Accessed 22 Jul. 2022 (2022)

9. Tang, J.: Understanding the Telework Experience of People with Disabilities. Proceedings of the ACM on Human-Computer Interaction. 5:1–27 (2021) https://doi.org/10.1145/3449104
10. Goldberg, E.: A Two-Year, 50-Million-Person Experiment in Changing How We Work. New York Times, March 10, 2022. https://www.nytimes.com/2022/03/10/business/remote-work-office-life.html. Accessed 24 Jul. 2022 (2022)
11. Boden, S.: Remote work is commonplace now, and workers with disabilities could benefit from the change. WESA Radio Pittsburgh, Apr. 3, 2022. https://www.wesa.fm/health-science-tech/2022-04-05/remote-work-is-commonplace-now-and-workers-with-disabilities-stand-to-benefit. Accessed 24 Jul. 2022 (2022)
12. Strickland, K.: Remote Work as a Reasonable Accommodation: Implications from the COVID-19 Pandemic. Harvard Civil Rights - Civil Liberties Law Review. https://harvardcrcl.org/remote-work-as-a-reasonable-accommodation-implications-from-the-covid-19-pandemic/. Accessed 24 Jul. 2022 (2021)
13. Hsu, A.: NPR article: The idea of working in the office, all day, every day? No thanks, say workers. https://www.npr.org/2022/06/05/1102744672/remote-work-from-home-return-to-office-covid-pandemic-workers-apple-google. Accessed 16 Jul. 2022 (2022)
14. Christian, A.: BBC News article: Why hybrid work is emotionally exhausting. https://www.bbc.com/worklife/article/20220120-why-hybrid-work-is-emotionally-exhausting, Accessed 16 Jul. 2022 (2022)
15. Christian, A.: BBC News article: The employee surveillance that fuels worker distrust. https://www.bbc.com/worklife/article/20220621-the-employee-surveillance-that-fuels-worker-distrust. Accessed 16 Jul. 2022 (2022)